\documentstyle[12pt,fleqn,epsfig]{article}
\newlength{\dinwidth}
\newlength{\dinmargin}
\newcommand{\resection}[1]{\setcounter{equation}{0}\section{#1}}

\newcommand{\f}[2]{\frac{#1}{#2}}

\begin{document}                   
\def\ra{\rightarrow}
\def\dt{\mbox {\boldmath $\Delta$}}
\def\pmu{p_\mu}
\def\ds{\displaystyle}
\def\tb{\bar{\tau}}
\def\ep{\varepsilon}
\def\pbmu{\bar{p}_{\mu}}
\def\pbnu{\bar{p}_{\nu}}
\def\pb{\bar{p}}
\def\k{{\bf k}}
\def\as{\alpha_s}
\def\q{{\bf q}}
\def\FF{{\cal F}}
\def\ss{\sigma}
\def\ssh{\hat{\sigma}}
\def\bks{\!\!\!\!\!\!\!\!\!}
\def\th{\hat{t}}
\def\uh{\hat{u}}
\def\sh{\hat{s}}
\def\o{\omega}
\def\g{\gamma}
\def\G{\Gamma}
\def\sst{\scriptscriptstyle}
\newcommand{\be}{\begin{equation}}
\newcommand{\ee}{\end{equation}}
\newcommand{\bea}{\begin{eqnarray}}
\newcommand{\eea}{\end{eqnarray}}
\newcommand{\nn}{\nonumber}

 \begin{flushright}DFF 250/6/96\\ 
 HEP-PH 9606427 
 \end{flushright}
 \vspace*{.5 cm}
 \begin{center}
 {\Large \bf Non-abelian $q\bar{q}$ contributions to small-$x$\\
 \vspace*{.6 cm} anomalous dimensions$^*$}\\
 \vspace*{1 cm}
 {\large G. Camici and M. Ciafaloni\\ \vspace*{3 mm}
 {\em Dipartimento di Fisica, Universit\`a di Firenze\\ \vspace*{3 mm}
 and INFN, Sezione di Firenze}}
 \end{center}

 \begin{abstract}
By using $\k$-factorization, we derive resummation 
formulas for the non-abelian $q\bar{q}$ contributions 
to both heavy flavour production by gluon fusion, and to the 
next-to-leading BFKL 
kernel. By combining this result with previous ones by Fadin et al. on 
the virtual terms, we also compute in closed form the complete $q\bar{q}$ 
contribution to the gluon anomalous dimension in the $Q_0$-scheme.
We find that $q\bar{q}$ resummation effects are important for heavy 
flavour production, but are instead small in the anomalous dimension 
eigenvalues, because of a cancellation between abelian and non abelian 
contributions.
 \end{abstract}
 \begin{center}
 \end{center}
 \vspace*{1.5 cm}
 $~^*$Work supported in part by the EC program CHRX-CT93-0357 and 
               by MURST (Italy).
  \cleardoublepage

\resection{Introduction}

Hard processes in the small-$x$ regime $s\gg Q^2\gg\Lambda^2$ play 
an increasingly important role at present colliders \cite{1}
and are characterized by 
the fact that the effective QCD coupling constant $\as(Q^2)\log(1/x)$ is  
sizeable and the corresponding perturbative contributions need to be 
resummed.

Such resummations are performed, at leading log level, on the basis of the BFKL 
equation \cite{2} for the anomalous dimensions, and of the 
$\k$-factorization formulae [3-6] for the coefficient functions.
Particular attention has been devoted in the literature to heavy quark 
production 
processes \cite{3,6} and to the (abelian) light quark contributions to
the $qg$ and $qq$ entries of the 
anomalous dimension matrix\cite{5}. Furthermore, an impressive program of 
evaluation of next-to-leading (NL) kernels of the BFKL equation is under way
[7-10].

In this note, we focus our attention on the (non-abelian) $q\bar{q}$ 
contribution to the $\k$-factorization program, which is relevant in two ways:
 firstly, as  a hard subprocess in Drell-Yan heavy-flavour production 
and secondly, in the massless quark limit, as an important NL contribution to 
the {\em gluon} anomalous dimension. 
 We shall then derive QCD resummation formulas in both cases and 
in particular we shall obtain in closed form all NL terms 
of the form $\as N_f(\as C_A\log(1/x))^n$, contributing to the gluon 
anomalous dimension in the so-called $Q_0$-scheme \cite{11,12}.

The evaluation of such contributions is particularly important for structure 
functions. Since the gluon couples to the proton only through 
$q\bar{q}$ states, the NL anomalous dimension $\g_{qg}$ (determined 
essentially by the abelian $q\bar{q}$ kernel) contributes to scaling violations 
at the same level as $\g_{gg}$.
It is then important to perform a complete calculation of NL $q\bar{q}$ 
contributions (in particular, the non-abelian one), in order to check 
the pattern described above, and to reduce the factorization scheme 
dependence of the theoretical estimates of scaling violations at HERA 
\cite{13,14}.

\resection{$Q\bar{Q}$ Hadroproduction}

Let us start considering the cross-section for the hadroproduction of a (heavy) 
quark-antiquark pair, which at high energies is dominated by the gluon fusion 
process in Fig. 1. 
This contribution may be expressed in the $\k$-factorized form \cite{3}:
\be
M^2\ss=\int\f{dz_1}{z_1}\f{dz_2}{z_2}d^2\k_1d^2\k_2\ssh(\k_1,\k_2,M^2,z_1z_2s)
\FF^{(1)}(z_1,\k_1)\FF^{(2)}(z_2,\k_2),
\ee
where $\FF$ denotes the unintegrated gluon density in the hadron
and $\ssh$ the high-energy projection of the (Regge) gluon fusion process 
$g(\k_1)g(\k_2)\ra Q\bar{Q}$. 

Since the evolution of the structure functions 
is simpler in the space of $z$ and $\k$ moments it is useful to express the 
high energy factorization formula (2.1) in terms of the double-Mellin 
transformed structure functions
\be
\FF_{\o}^{(i)}(\g)=\int^1_0dzz^{\o-1}\int_0^\infty\f{d^2\k}{\pi\k^2}
\left(\f{\k^2}{Q_0^2}\right)^\g\FF^{(i)}(z,\k)
\ee
and hard cross-section coefficient function \footnotemark
\footnotetext{This function is related by the change of normalization 
$h=\f{4\as}{\pi}\g_1\g_2 H$ to the $h$-functions of Refs \cite{3,5,6}}
\be
\left (\f{\as}{\pi}\right)H_\o(\g_1,\g_2)=
\int_0^\infty\f{d^2\k_1}{\k_1^2}\f{d^2\k_2}{\k_2^2}
\left(\f{\k_1^2}{M^2}\right)^{\g_1}\left(\f{\k_2^2}{M^2}\right)^{\g_2}
\int_0^\infty\f{ds}{s}\left(\f{M^2}{s}\right)^\o \ssh(\k_1,\k_2,M^2,s),
\ee
so that it takes the form:
\be
M^2\ss\!\left(\f{M^2}{s}\right)\!=\!\left(\f{\as}{\pi}\right)\!\!
\int_{\f{1}{2}-i\infty}^{\f{1}{2}+i\infty}\!\!i\f{d\g_1d\g_2d\o}{(2\pi)^3}
\left(\f{s}{M^2}\right)^\o
\left(\f{M^2}{Q_0^2}\right)^{\g_1+\g_2}\!\!\FF_{\o}^{(1)}(\g_1)
\FF_{\o}^{(2)}(\g_2)H_\o(\g_1,\g_2).
\ee 
$~$
\indent For large enough $M^2$, the $\g$ integrals will be dominated by 
the BFKL anomalous dimension, $\g_1\simeq\g_2\simeq\g_L(\bar{\as}/\o)$. This  
means that the high-energy resummation effects 
on the cross section we are considering are embodied in the 
$\g_1,~\g_2$-dependence of the hard coefficient function 
$H_\o(\g_1,\g_2)$, which is precisely what we wish to compute. 

The squared matrix element for the process under consideration 
was computed 
in Ref. \cite{3} (see also Ref. \cite{8}), 
and contains two terms, with colour factors $C_F$
and $C_A$, that we shall call respectively the abelian  
and the non abelian contributions.

The abelian contribution $H^a$ is known \cite{3} 
and provides the quark anomalous 
dimension \cite{5}.
Let us then concentrate on the non abelian one $H^{na}$.

First, we rewrite the squared matrix element of Ref. \cite{3} in terms of the 
Sudakov parametrization for the exchanged gluons' momenta
\be
k_1^\mu\simeq z_1p_1^\mu+\k_1^\mu,~~~~~k_2^\mu\simeq z_2p_2^\mu+\k_2^\mu,
\ee
and for the momentum transfer
\be
\Delta^\mu=z_1x_1p_1^\mu-z_2x_2p_2^\mu+\dt^\mu,
\ee
where $p_1$, $p_2$ denote the (light-like) momenta of the incoming hadrons.

By using explicit expressions for the invariants
\be
\sh = (k_1+k_2)^2,~~~~\th = \Delta^2,~~~~\uh = (k_1-k_2-\Delta)^2,
~~~~\nu=\sh+(\k_1+\k_2)^2,
\ee
the non abelian squared matrix element of Ref. \cite{3} can be rewritten 
in the form
\bea
& &\bks A^{na}=\pi^2\as\left[\f{-1}{(M^2-\th)(M^2-\uh)}+\f{1}{\sh}
\left(\f{1}{M^2-\uh}-\f{1}{M^2-\th}\right)(1-x_1-x_2)+\f{2}{\nu\sh}+\right.\nn\\
& &\bks+\f{2}{\k_1^2\k_2^2}
\left(\f{1}{2}-\f{(1-x_1)(1-x_2)\nu}{M^2-\th}+\f{\nu}{2}(1-x_1-x_2)-\right.\nn\\
& &\bks-\left.\f{\k_1^2(1-x_2)+\k_2^2(1-x_1)-\k_1\cdot\k_2+
\dt\cdot(\k_2-\k_2)}{\sh}\right)\times\nn\\
& &\bks\left(\f{1}{2}-\f{x_1x_2\nu}{M^2-\uh}-\f{\nu}{2}(1-x_1-x_2)+\right.
\left.\left.\f{\k_1^2(1-x_2)+
\k_2^2(1-x_1)-\k_1\cdot\k_2+\dt\cdot(\k_2-\k_2)}{\sh}\right)\right]\nn\\
\eea

By then integrating over the phase space and performing the moments of Eq. (2.3)
we obtain the non abelian coefficient function $H^{na}_\o(\g_1,\g_2)$
as follows
\bea
& &\bks \f{\as}{\pi}H_\o^{na}(\g_1,\g_2)=
\f{1}{8\pi^4}\int\!\!\f{d\nu}{\nu^{2-\o}}\f{d^2\k_1}{\pi\k_1^2}
\f{d^2\k_2}{\pi\k_2^2}\f{d^2\dt}{\pi}dx_1dx_2\delta(x_2(1-x_1)\nu-(\k_1-\dt)^2
-M^2)\times\nn\\
& &\bks\times\delta(x_1(1-x_2)\nu-(\k_2+\dt)^2-M^2)
\left(\f{\k_1^2}{M^2}\right)^{\g_1}
\left(\f{\k_2^2}{M^2}\right)^{\g_2}
A^{na}(\k_1,\k_2,\dt,x_1,x_2,\nu,M^2)
\eea
This expression can be evaluated analytically (at least in the limit 
$\o=0$, relevant at high 
energies) with a careful choice of the order of the integrations.
We find it convenient to start eliminating the variable $\nu$ by 
integration of one mass-shell delta function, and then computing 
the $\dt$-integral with the aid of a Feynman parametrization of denominators.
The remaining integrals, first over the transverse momenta $\k_1$, $\k_2$ and 
then over the  
Feynman parameters and longitudinal momentum fractions, can be 
evaluated in terms of Gamma and Beta 
functions, and give, after some algebra, the result
\bea
& &\bks H^{na}(\g_1,\g_2)=\nn\\
& &\bks =\f{C_A}{2}\f{\as}{\pi}\left[
\G(1-\g_1-\g_2)\G(\g_1)\G(\g_2)\left(\f{B(1-\g_1,1-\g_1)B(1-\g_2,1-\g_2)}{4}+
\right.\right. \nn \\
& &\bks\left.+\f{B(1-\g_1,2-\g_1)B(1-\g_2,2-\g_2)}{(3-2\g_1)(3-2\g_2)}
(1+(1-\g_1)(1-\g_2))\right)+\nn\\
& &\bks+B(\g_1,1-\g_1)B(\g_2,1-\g_2)
\left(\f{\G(2-\g_1-\g_2)}{\G(4-2\g_1-2\g_2)}-
\f{\G(3-\g_1-\g_2)}{\G(6-2\g_1-2\g_2)}\right)+\nn\\
& &\bks+2\!\!\left.\f{B(\g_1,1-\g_1-\g_2)B(\g_2,1-\g_1-\g_2)
B(3-\g_1-\g_2,3-\g_1-\g_2)}{(1-\g_1-\g_2)}\!
\right].
\eea
The details of the calculation are reported elsewhere \cite{15}.

Let us note the triple pole singularity at $\g_1+\g_2=1$
 of the last term of Eq. (2.10), which 
is related to the collinear-singular behaviour of the 
partonic cross section in the massless limit.
In fact, if 
$M^2\ll(\k_1+\k_2)^2\equiv \q^2\ll\k_1^2\simeq\k_2^2$, the hard cross 
section approaches the singular limit
\bea
& &\bks\!\ssh^s_{\!\o=0}(\k_1,\k_2,M^2)\!\! =\!\!
\bar{\as}\f{\as}{\pi}\!\int_0^1\!\!\!dx_1\f{x_1(1\!-\!x_1)}{\q^2}\log\left(\!
1+\!\f{x_1(1\!-\!x_1)\q^2}{M^2}\right)\!
\simeq\!\bar{\as}\f{\as}{6\pi}\left[\log\f{\q^2}{M^2}-
\f{5}{3}\right]\f{1}{\q^2}\nn\\
& &
~~~~~~~~~~~~~~~~~~~~~~~~~~~~~~~~~~~~~~~~~~~~~~~~~~~~~~
\left(\bar{\as}=\f{\as C_A}{\pi}\right),
\eea
which, after taking the $\k_i^2$-moments, yields precisely a triple pole.
The factor in front of the logarithm is related to the $g\ra q\bar{q}$ GLAP
splitting function \cite{3,6}.

Such singular behaviour was already discussed in Ref. \cite{3}
and yields an interesting enhancement of the cross section when 
$\g_1$ and $\g_2$ approach the BFKL saturating value $\g=1/2$.
Here we further notice that, after combination with virtual terms in 
a BFKL kernel, the behaviour (2.11) is responsible for the $q\bar{q}$ 
contribution to the running coupling constant, as we shall see in the following.

Indeed, the same cross section 
$\ssh$ occurs, in the massless limit, as a light quark contribution 
to the BFKL kernel, given by the inverse of Eq. (2.3), i.e., by
\be
\f{\pi}{\as M^2}\ssh_{\o=0}\left(\k_1,\k_2,M^2\right)=
\f{1}{M^2}\int_{\f{1}{2}-i\infty}^{\f{1}{2}+i\infty}
\f{d\g_1 d\g_2}{(2\pi)^2}H_0(\g_1,\g_2)\left(\f{\k_1^2}{M^2}\right)^{-\g_1}
\left(\f{\k_2^2}{M^2}\right)^{-\g_2}.
\ee

After subtraction of the singular contribution of Eq. (2.11), the eigenvalue  
of the finite part of the kernel is directly obtained by the massless limit of 
Eq. (2.12), which singles out 
the residue at the (single) pole at $\g_1+\g_2=1$, as follows
\bea
& &H_{real}^{na}(\g)=\lim_{\g_2\ra 1-\g}\left(H^{na}(\g,\g_2)-H^{na}_s(\g,\g_2)
\right)(1-\g-\g_2)\nn\\
& &=\left(\f{\as}{\pi}\right)\f{C_A}{2}
\left[\G^2(\g)\G^2(1-\g)\left(\f{1}{3}-
(2+3\g(1-\g)\f{\G(2-\g)\G(1+\g)}{\G(4-2\g)\G(2+2\g)}\right)\right],
\eea
for each quark flavour.

Note that the various terms of Eq. (2.13) show spurious double poles 
at $\g=0$ and $\g=1$, which would signal the presence of a collinear 
singularity in the $t$-channel. They cancel however in the sum, as they 
should because the real emission non-abelian diagrams do not contribute to 
the t-channel singularity.

\resection{Next-to-leading gluon anomalous dimension}

Let us now come to the important issue of NL contributions to the flavour
singlet anomalous dimension matrix $\g_{ab}(\as)$ ($a,b=q,g$).
As is known, the gluon entries start at leading level
\be
\g_{gg}=\g_{L}\left(\f{\bar{\as}}{\o}\right)+\as\g_{NL}\left(\f{\bar{\as}}
{\o}\right)+....,~~~~~~~\g_{gq}=\f{C_F}{C_A}\g_L(1+O(\as)),
\ee
while the quark  entries \cite{5} start at NL level and, in the 
$Q_0$-scheme \cite{12} are given by
\be
\g_{qg}=N_f\g_L^2H_2(\g_L),~~~~~\g_{qq}=\f{C_F}{C_A}(\g_{qg}-\f{N_f\as}{3\pi}).
\ee
Here $H_2$ is the $\k$-factorization vertex measured by the $F_2$ 
structure function and, similarly to Eq. (2.13), is given by the 
abelian $q\bar{q}$
contribution as \cite{5}
\bea
& &\bks H_2(\g)=C_F^{-1}\lim_{\g_2\ra 1-\g} H_o^{ab}(\g,\g_2)(1-\g-\g_2)=\nn\\
& &\bks \f{\as}{\pi}
\f{\G^2(\g)\G^2(1-\g)\G(2-\g)\G(1+\g)}{\G(4-2\g)\G(2+2\g)}
\left(2+3\g(1-\g)\right).
\eea
The NL terms of $\g_{gg}$ occur in the larger of the anomalous 
dimension eigenvalues
\be
\g_+\simeq\g_{gg}+\f{C_F}{C_A}\g_{qg},~~~~\g_-\simeq -\f{C_FN_f\as}{3\pi C_A}
\ee
and require a full discussion of the BFKL kernel
\be
K_\o(\as,\k_1,\k_2)=\f{\bar{\as}(\k_1^2)}{\o}
\left(K_0(\k_1,\k_2)+\as K_1(\k_1,\k_2)\right).
\ee
Here $K_0$ is the leading kernel, whose $\g$-dependent eigenvalue \cite{1} 
\be
\chi_0(\g)=2\psi(1)-\psi(\g)-\psi(1-\g) 
\ee
determines the anomalous dimension $\g_L(\bar{\as}/\o)$
\be
1=\f{\bar{\as}}{\o}\chi_0(\g_L),~~~~\g_L=\f{\bar{\as}}{\o}+
2\zeta(3)\left(\f{\bar{\as}}{\o}\right)^2+
...~~~~~~~~~~.
\ee

Furthermore, both $K_0$ and the NL kernel $K_1$ are scale-invariant,
because the non-invariant NL part has been factored out in Eq. (3.5) 
inside the running coupling constant at the scale $\k_1^2$ \cite{16,17}

\be
\as(\k_1^2)=\as(\mu^2)-b\as^2(\mu^2)\log\f{\k_1^2}{\mu^2}+...,~~~~~
\mu=O(Q).
\ee
We shall compute the NL contributions (which are factorization scheme 
dependent \cite{11,18}), in the so-called $Q_0$-scheme of Ref. \cite{12}, 
further discussed in a forthcoming paper \cite{15}.
In this scheme we define the quark sea density as in the
physical DIS scheme \cite{19}, i.e. by the structure function $F_2$,
similarly to the parton model.

Instead, the gluon density is defined by the procedure of 
$\k$-factorization itself, which provides a gauge invariant off shell
continuation, so that the initial gluon virtuality can be fixed at $Q_0$.  
Therefore, gluon states carry a high-energy polarization which 
in general does not match the collinear definition at higher orders.
For this reason, the expression of $\g_{qg}$ in Eq. (3.2) 
does not carry the R-factors of the ${MS}$ type schemes \cite{11,5},
and a leading similarity transformation \cite{12} is needed to relate 
the former scheme to the latter.

The key advantage of the $Q_0$-scheme is that the anomalous dimension
$\g_+$ of Eq. (3.4) is 
computed from the eigenvalue of the scale invariant kernel
$K_0+\as K_1$, directly in four dimensions, as follows:

\be
1=\f{\bar{\as}}{\o}\left(\chi_0(\g_+)+\as\chi_1(\g_+)\right),
~~~~~~~~~~~~~~~~\bar{\as}=\bar{\as}(\mu^2).
\ee

By expanding and using Eq. (3.7), we obtain the NL
contribution to $\g_+$
\be
\g_+^{NL}\left(\f{\bar{\as}}{\o}\right)=-\as
\f{\chi_1\left(\g_L(\bar{\as}/\o)\right)}
{\chi_0^\prime\left(\g_L(\bar{\as}/\o)\right)}=
\g_{gg}^{NL}+\f{C_F}{C_A}\g_{qg},
\ee
which contains, in particular, the $q\bar{q}$ contribution.

In order to compute $\chi_1^{na}$, the non-abelian part of 
$\chi_1^{q\bar{q}}$, we need to combine the singular real emission 
contribution in Eq. (2.11) with the virtual corrections, and 
then to add the resulting finite term to the nonsingular real emission 
part in Eq. (2.13).

The virtual $q\bar{q}$ contribution to the one gluon emission kernel 
was computed in Ref.\cite{9} and, after azimuthal averaging, takes the form
\be
\f{\as N_f}{6\pi}
\left(\f{1}{\q^2}\log\f{M^2}{\mu^2}-
\f{1}{\k_1^2-\k_2^2}\log\f{\k_1^2}{\k_2^2}\right),~~~~~~~~~~~~~~~~~~~
(\q=\k_1+\k_2)
\ee
in which $\mu^2=O(Q^2)$ is the factorization scale, and $M$ is the (small)
quark mass.
By combining Eq. (3.11) with the singular $q\bar{q}$ emission part in 
Eq. (2.11) the $M^2$-dependence cancels out and we get
\be
\f{\as N_f}{6\pi}\left[\left(\log\f{\k_1^2}{\mu^2}+
\log\f{\q^2}{\k_1^2}-\f{5}{3}\right)\f{1}{\q^2}-
\f{1}{\k_1^2-\k_2^2}\log\f{\k_1^2}{\k_2^2}\right].
\ee

A similar calculation can be performed \cite{15} directly at $M=0$ and 
$D=4+2\ep$ dimensions, to yield the $\ep$-dependent kernel
\be
\f{\as N_f}{6\pi}\left[\f{1}{\ep}\f{\G^2(1+\ep)}{\G(1+2\ep)}
\left(\f{\q^2}{\mu^2}\right)^\ep\f{1-\f{5}{3}\ep+\f{28}{9}\ep^2+O(\ep^3)}{\q^2} 
-\f{1}{\ep}\f{1}{\q^2}-\f{1}{\k_1^2-\k_2^2}\log\f{\k_1^2}{\k_2^2}\right].
\ee

In order to regularize the $\q^2=0$ singularity in Eq. (3.12) and (3.13), 
one should combine the above results with the purely virtual term, computed in 
Ref.\cite{10}. This can be done in dimensional regularization by means of 
Eq. (3.13), and the result \cite{15} can be read off directly in four 
dimensions as follows: the logarithmic term at scale $\k_1$ in Eq. (3.12) 
is absorbed in the running coupling factor of Eq. (3.5) and the remaining 
(regularized) collinear part becomes
\be
\as(K_1^{(V)}+K_1^{na,s})=
\f{\as N_f}{6\pi}\left[\left.\left(\log\f{\q^2}{\k_1^2}-\f{5}{3}\right)
\f{1}{\q^2}
\right|_R
-\f{1}{\k_1^2-\k_2^2}\log\f{\k_1^2}{\k_2^2}\right]
\ee
where we have introduced the notation
\be
f(\k_1,\q)\left.\right|_R=f(\k_1,\q)-\delta^2(\q)\int_{\lambda^2}^{\k_1^2}
d^2\q f(\k_1,\q)
\ee
and $\q^2>\lambda^2$ is understood in all $\q$-integrations.
It is now straightforward to diagonalize the kernel (3.14) and to combine it 
with the finite real emission part in Eq. (2.13) to get the complete 
non-abelian eigenvalue expressed in terms of $\chi_0(\g)$ of Eq. (3.6)
and of $\psi(\g)$
\bea
& &\bks\as\chi_1^{(na)}(\g)=
\f{N_f\as}{6\pi}\left[
\f{1}{2}(\chi_0^\prime+\chi_o^2)-\f{5}{3}\chi_0-\left(\psi^\prime(\g)+
\psi^\prime(1-\g)\right)+\right.\nn\\
& &\left.\bks\left(\f{\pi}{\sin\pi\g}\right)^2
\left(1-3\cos(\pi\g)\f{1+\f{3}{2}\g(1-\g)}{(3-2\g)(1-2\g)(1+2\g)}\right)\right],
\eea
together with its abelian counterpart of Eq. (3.3):
\be
\as\chi_1^{(a)}=\f{C_F}{C_A}\f{\as N_F}{\pi}
\left(\f{\pi}{\sin\pi\g}\right)^2
\cos(\pi\g)\f{1+\f{3}{2}\g(1-\g)}{(3-2\g)(1-2\g)(1+2\g)}.
\ee

Note that both eigenvalues are symmetrical for $\g\ra 1-\g$, apart 
from the first term in square brackets of Eq. (3.16) ($\sim\chi_0^\prime$) 
which comes from the fact
that we have factorized the running coupling effects in the upper mass $\k_1^2$.

In order to extract from Eqs. (3.16) and (3.17) the anomalous dimension 
eigenvalue $\g_+$ we have to use Eq. (3.10), i.e., we have to divide it by 
$\chi_0^\prime$
Since $\chi_0^\prime(\g)$$\simeq-1/\g^2+O(\g)$ for small $\g$, the 
one-loop and two-loop results in the DIS scheme 
are read off from the small $\g$ behaviour of Eqs. (3.16), (3.17), that is
\be
\as\chi_1^{(a)}(\g)\stackrel{\displaystyle \simeq}{\sst \g\ll 1}\!
\f{C_F}{C_A}\f{\as N_f}{3\pi}\!\!\left(\f{1}{\g^2}+\f{13}{6\g}+O(1)\!\!\right),
~~~
\as\chi_1^{(na)}\stackrel{\displaystyle \simeq}{\sst\g\ll 1}\!
\f{\as N_f}{6\pi}\!\!\left(-\f{1}{\g^2}-\f{23}{6\g}+O(1)\!\!\right)\!.
\ee

In fact, by using Eq. (3.10) we check directly the known \cite{20} low-order 
expressions in the DIS scheme
\be
\g_{qg}=\f{N_f\as}{3\pi}\left(1+\f{13}{6}\f{\bar{\as}}{\o}+...\right),
~~~~~~~
\g_{gg}^{\sst NL}=-\f{N_f\as}{6\pi}\left(1+\f{23}{6}\f{\bar{\as}}{\o}+
...\right).
\ee

Furthermore, Eqs. (3.16), (3.17) 
provide an all-order resummation in the $Q_0$-scheme.
Since $\g_{qg}$ is known from Eq. (3.2), and $-\chi_0^\prime\g^2$
$=1+O(\g^3)$, we find that $\chi_1^a$ is not perfectly cancelled by 
$(C_F/C_A)\g_{qg}$, and therefore $\g_{gg}^{NL}$ takes also contributions 
of type $\f{N_{F}C_F\as}{\pi C_A}\left(\f{\bar{\as}}{\o}\right)^n$ from 
three loops on. Finally, the terms $\f{N_F\as}{\pi}
\left(\f{\bar{\as}}{\o}\right)^n$ are resummed by $\chi_1^{na}$, which is 
typically  of negative sign, as it appears already at low order in Eq. (3.19).

The total $q\bar{q}$ eigenvalue $\chi_1^{q\bar{q}}(\g)$
$=\chi_1^a(\g)$$+\chi_1^{na}(\g)$ is plotted against  $\g$ in Fig. 2. 
It is apparent that a cancellation between abelian and non abelian terms is at 
work, and that resummation effects are small, in their sum, even for 
$\g$ values which approach the saturation limit $\g=1/2$.
Only for $\g\ra 1$ is the abelian term dominant, because of its stronger 
$\g=1$ singularity.

This remark is made more significant by the observation that the 
scheme dependence of $\g_+$ is a simple one.
For instance, the transformation from the DIS-$Q_0$ scheme to the 
DIS-$\overline{MS}$  scheme, involves a scale change by a factor  
$R(\g_L(\as(t)))$ \cite{11,12}, under which $\g_+$ changes by running coupling 
effects only, i.e.,
\be
\tilde{\g}_+=\g_+-\f{\dot{R}}{R}.
\ee

We tentatively conclude that $q\bar{q}$ resummation effects 
in scaling violations are 
small in a combination proportional to 
$\FF_q+(C_A/C_F)\FF_g$ which isolates the $\g_+$ 
eigenvalue, and is accessible experimentally through proper combinations 
of $F_2$ and $F_L $\cite{15,18}, when the latter will be measured. 
It is important to check 
this conclusion against the purely gluonic NL terms, which are still under 
investigation \cite{7,8}.

A final comment is needed for the running coupling effects (to which the 
$q\bar{q}$ channel contributes)
which remain an important NL feature of the BFKL equation.
We have decided to factorize the $\as$ dependence at the upper scale 
$\k_1^2$ \cite{17} in Eq. (3.5). It can be shown \cite{15} that, in the regime 
$\k_1^2\gg Q_0^2\gg\Lambda^2$, with $\bar{\as}(Q_0^2)/\o <1$, 
 this dependence is equivalent to a renormalization-group evolution due to the 
anomalous dimension $\g_+$ of Eq. (3.9), apart from a calculable 
coefficient factor, common to quark the gluon in the $Q_0$-scheme, 
given by the expression
\be
\left[\g_L\left(\f{\bar{\as}}{\o}\right)\sqrt{-\chi_0^\prime(\g_L)}\right]^{-1}.
\ee

Since the latter differs from 1 only from 3 loops on, the choice we have 
made is good at perturbative level. The factor (3.21) can be 
reinterpreted in terms of an 
additional contribution to the gluon anomalous dimension, 
along the lines of Eq. (3.20).

To sum up, in this paper we have resummed in closed form the non abelian
$q\bar{q}$ contributions to both heavy flavour production and to the small-$x$ 
anomalous dimensions.
While non abelian resummation effects are quite important in the former case, 
they nearly cancel with the abelian ones in the latter, due to virtual 
contributions in the anomalous dimension eigenvalue.

The most significant NL effects (apart from the running coupling) seem to 
remain the rather large scaling violations due to $\g_{qg}$ \cite{13}, which 
somehow represent a leakage from the gluon density to the quark sea, 
which unfortunately depends on how the gluon is defined, even in the class 
of DIS-schemes.
It is therefore important to measure a quantity independent of $F_2$, 
for instance $F_L$, in order to disentangle the gluon density from the quark 
sea and/or to look 
for combinations with smaller scaling violations, as is suggested by the 
result on $\g_+$ of the present work.

\newpage

\begin{figure}[htb]
\vspace*{-0.8 cm}
\centerline{\psfig{figure=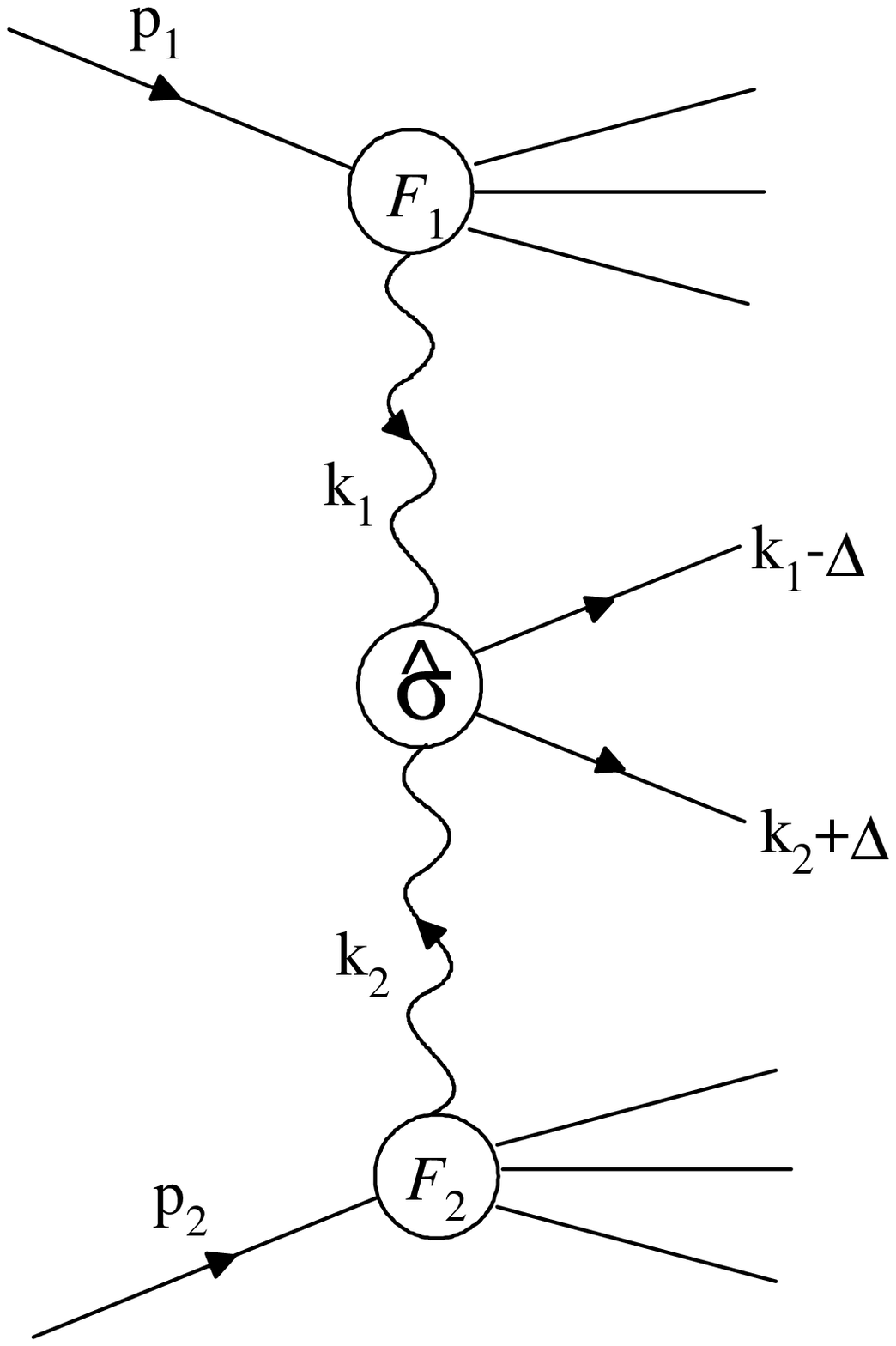}}
\vspace*{-10 cm}
\flushleft{Fig. 1: Kinematics of the $Q\bar{Q}$ hadroproduction.
Wavy lines denote \\ (Regge) gluon exchanges.}
\end{figure}

\newpage

\begin{figure}[htb]
\vspace*{-2.8 cm}
\centerline{\psfig{figure=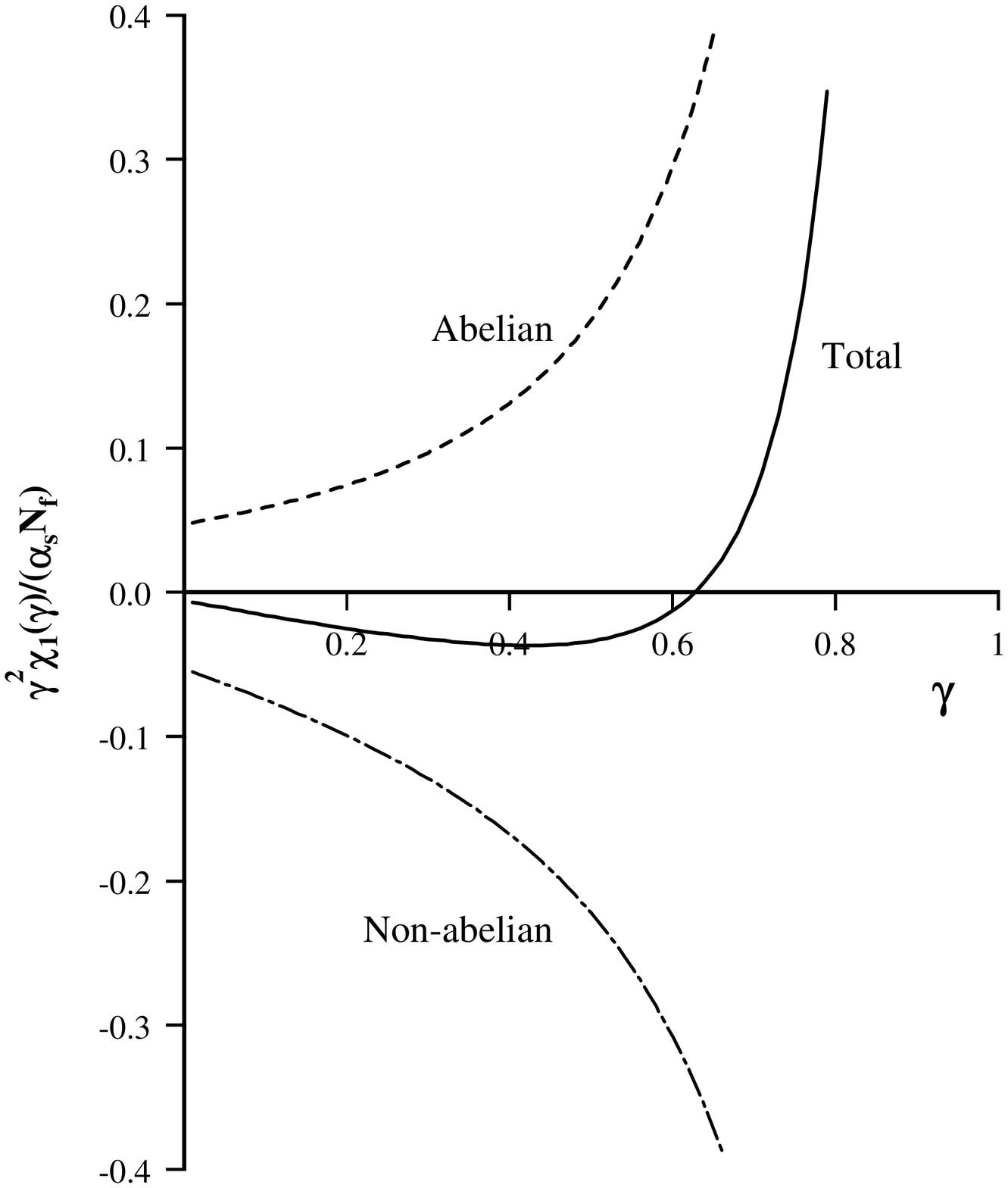}}
\vspace*{-5 cm}
\flushleft{Fig. 2: $q\bar{q}$ contributions to the BFKL eigenvalue as  
function of the anomalous \\ dimension variable $\g$.}
\end{figure}

\end{document}